\documentclass[twocolumn,aps,showpacs,prb,tightenlines,amsmath,amssymb,superscriptaddress]{revtex4}
\usepackage{graphicx}
\usepackage{amssymb}
\usepackage{dcolumn}
\usepackage{amsmath}

\usepackage{bm}

\usepackage{colordvi}

\begin{document}

\title{$L$-valley electron $g$ factor in bulk GaAs and AlAs}
\author{K. Shen}
\affiliation{Hefei National Laboratory for Physical Sciences at
Microscale, University of Science and Technology of China, Hefei,
Anhui, 230026, China}
\author{M. Q. Weng}
\email{weng@ustc.edu.cn.}
\affiliation{Department of Physics,
University of Science and Technology of China, Hefei,
Anhui, 230026, China}
\author{M. W. Wu}
\email{mwwu@ustc.edu.cn.}
\affiliation{Hefei National Laboratory for Physical Sciences at
Microscale,
University of Science and Technology of China, Hefei,
Anhui, 230026, China}
\affiliation{Department of Physics,
University of Science and Technology of China, Hefei,
Anhui, 230026, China}

\date{\today}

\begin{abstract}
We study the Land\'e $g$-factor of conduction electrons in the
$L$-valley of bulk GaAs and AlAs by using a three-band
$\mathbf{k}\cdot\mathbf{p}$ model together with the tight-binding model.
We find that the $L$-valley $g$-factor
is highly anisotropic, and can be characterized by two components,
$g_{\perp}$ and $g_{\|}$. $g_{\perp}$ is close to the free electron
Land\'e factor but $g_{\|}$ is strongly affected by the remote bands.
The contribution from remote bands on $g_{\|}$ depends on how the
remote bands are treated. However, when the magnetic field is in
the Voigt configuration,
which is widely used in the experiments, different models give almost
identical $g$-factor.
\end{abstract}


\maketitle

The knowledge to electron $g$-factor of semiconductors is of
fundamental importance to investigate
the spin-related problems. In the past
decades, a lot of investigations have been carried out to understand the
material and structure dependence of the
$g$-factor.\cite{ando,winkler,hannak,jeune,malinowski,weisbuch,kikkawa}
Experimentally, $g$-factor can be measured through the measurement of Lamour frequency by
the techniques such as time resolved Faraday/Kerr
rotation,\cite{malinowski,kikkawa} or through
the measurement of the Zeeman splitting by electron spin
resonance.\cite{weisbuch} Theoretical studies of the $g$-factor are carried out
through various band structure calculations, such as
$\mathbf{k}\cdot\mathbf{p}$ model, tight-binding model and the {\it
ab initio} calculation.\cite{chadi,jancu,cardona} For zinc blende materials,
 most of the previous studies
focus on the $g$-factors at $\Gamma$ point. As recent investigations
have been extended to the spin dynamics far away from
equilibrium,\cite{fu,saikin,jancu,zhang,jancu2} the knowledge of $g$
factor of higher valleys becomes important.  For III-V semiconductors, the
$g$-factor in the $X$-valley is believed to be close to the Land\'e
factor of free electron, $g_0$, regardless of the materials due to the
large band gap and the vanishing of the spin splitting at the
$X$-point.\cite{baron,sirenko} By using two-band
$\mathbf{k}\cdot\mathbf{p}$ Kane-like model,\cite{kane,baron} it is
shown that the $g$-factor in the vicinity of the $L$-valley of IV
semiconductors is highly anisotropic and can be characterized by
$g_{\|}$ and $g_{\perp}$ which correspond to
the $g$-factor  along the directions parallel and
perpendicular to the $L$-axis (i.e., along $[111]$ direction for the
$[111]$ valley).  However, the investigations
of $g$-factor in the $L$-valley of III-V semiconductors are still to
be carried out, especially for GaAs (Ga$_{1-x}$Al$_{x}$As),
one of the most promising materials for realizing the spintronic
device.\cite{ivchenko,kiselev,pfeffer2,pfeffer3,leyva1,leyva2} In
this paper, we present the $g$-factor of the $L$-valley of GaAs and
AlAs by $\mathbf{k}\cdot\mathbf{p}$ band structure calculation. With
these two $g$-factors, one can further obtain the $g$-factor of
Ga$_{1-x}$Al$_x$As by linear interpolation on $x$, which is
widely used in practice to get the material parameters of the
semiconductor alloy.\cite{hermann,chadi,vurgaftman}

\begin{table*}[bth]
  \caption{Band structure parameters and $g$ factors at $L$ point for
 GaAs and AlAs. The rows labeled
 ``$\alpha$'' and ``$\beta$'' represent
    the results from three-band model by choosing the value of
    $\lambda$ to be corresponding value of
    $L$-point of Ge and the value at $\Gamma$-point.\cite{winkler}
    The rows with ``$\gamma$'' denote the results from
    two-band model.}
  \begin{center}
    \centering
    \begin{tabular}{{p{1.1cm}p{0.5cm}}*{12}{p{1.05cm}}}
      \\[-3pt]
    \hline
    \hline
    \\[-8pt]
    \ \
    &\ \ &$E_{L_{1c}}$& $E_{L_{3v}}$&$E_{L_{3c}}$ &$\delta$&$\delta^\prime$
    &$m_l$ & $m_t$ &$P$ & $P^\prime$ &$g_\perp$ & $g_\|$ &$g_x$\\
    \ \ &\ \ &
    (eV)&(eV)&(eV)&(eV)&(eV)&($m_0$)&($m_0$)&(eV\AA)&(eV\AA)&\ \ &\ \ \\[1pt]
    \hline
    \\[-7pt]
    \ \ &$\alpha$
    &$1.85^{a}$&$-1.20^{a}$&$5.47^{a}$&$0.22^{a}$&$0.08$&$1.9^{a,b}$&$0.075^{a}$
    &$12.5$&$ 2.75$&$2.03$ &$1.09$& $1.77$\\
    GaAs &$\beta$
    &$1.85^{a}$&$-1.20^{a}$&$5.47^{a}$&$0.22^{a}$&$0.08$&$1.9^{a,b}$&$0.075^{a}$
    &$13.5$&$6.14$&$2.03$ &$0.89$&$1.74$\\
    \ \ &$\gamma$
    &$1.85^{a}$&$-1.20^{a}$&$5.47^{a}$&$0.22^{a}$&$0.08$&$1.9^{a,b}$&$0.075^{a}$
    &$12.2$&$0$&$2.03$ &$1.14$&$1.78$\\
    \hline
    \\[-7pt]
    \ \ &$\alpha$
    &$2.581^c$&$-0.983^c$&$5.069^c$&$0.208^c$&$0.058$ &$1.9^{a,b}$&$0.096^{a,b}$
    &$11.9$ &$2.62$&$2.03$ &$1.41$&$1.85$\\
    AlAs &$\beta$
    &$2.581^c$&$-0.983^c$&$5.069^c$&$0.208^c$&$0.058$ &$1.9^{a,b}$&$0.096^{a,b}$
    &$15.0$&$8.00$&$2.03$&$0.93$&$1.74$\\
    \ \ &$\gamma$
    &$2.581^c$&$-0.983^c$&$5.069^c$&$0.208^c$&$0.058$ &$1.9^{a,b}$&$0.096^{a,b}$
    &$11.5$ &$0$&$2.03$ &$1.46$&$1.86$\\[1pt]
    \hline
    \hline
    \\[-3pt]
    \mbox{$^a$Ref.\ \onlinecite{madelung}; $^b$Ref.\ \onlinecite{klimeck}; $^c$Ref.\ \onlinecite{jancu3}}
  \end{tabular}
  \label{table1}
\end{center}
\end{table*}

\begin{figure}[bth]
\begin{center}
\includegraphics[height=5.5cm]{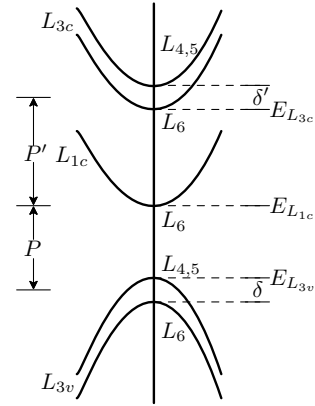}
\end{center}
\caption{Schematic of band structure near $L$ point for zinc-blende crystal.}
\label{fig1}
\end{figure}

Similar to the effective mass, the electron $g$-factor
of a specific band is affected by the remote bands.
The relevant bands in the vicinity of $L$-point
in our calculation are shown in
 Fig.~\ref{fig1} and are identified by their symmetries. The bands
with $L_3$ symmetry split to bands with $L_{4,5}$ and $L_6$
symmetries due to the spin-orbit coupling. $L_{1c}$ is
the lowest conduction band at which we target.
In addition to the top valence band
$L_{3v}$ included in the two-band Kane-like model,\cite{roth} we
further include
higher conduction bands $L_{3c}$ as the gap between $L_{3c}$ and
$L_{1c}$ bands is close to the gap between $L_{3v}$ and $L_{1c}$ in
bulk GaAs/AlAs.
In the calculation of $g_{\perp}$ we also include the
remote band $L_{2c}$.
The other remote bands are neglected in the calculation since
they are too far away from the conduction band to make any important
effect.\cite{madelung}
For the general information of the band structure and the
symmetry of zinc-blende materials, one can refer to the
literature, say Ref. \onlinecite{pyu}.

Follow the standard $\mathbf{k}\cdot\mathbf{p}$ calculation
procedure, the anisotropic $\bm g$ factor of $L_{1c}$ band can be
expressed as\cite{roth}
\begin{eqnarray}
\nonumber
&&\hspace{-1pc}{\bm g}=g_0{\bm 1}+\frac{2}{m_0i}\sum_{\mu\nu}\ \!\!
'\tfrac{1}{(E_{L_{1c}}-E_\mu)(E_{L_{1c}}-E_\nu)}
\{{\bm h}_{0\mu}
({\bm p}_{\mu\nu}\times{\bm p}_{\nu 0})\\ &&
+{\bm h}_{\mu\nu}({\bm p}_{0\mu}\times
{\bm p}_{\nu 0})+{\bm h}_{\nu
    0}({\bm p}_{\nu 0}\times{\bm p}_{\mu\nu})\}\ ,
\label{eq3}
\end{eqnarray}
where ${\bm p}_{\mu\nu}$ is inter-band momentum matrix element
between $\mu$ and $\nu$ bands. Noted that index ``$0$'' denotes
$L_{1c}$ band. ${\bm h}$ is the effective magnetic
field from the spin-orbit coupling. $g_0$ and $m_0$ are
the Land\'e factor and mass of free electron.
In the coordinate system defined by the principle axes of the constant
energy conduction band ellipsoid, {i.e.} $z$-axis parallels to
$\Gamma$-$L$ axis, the $g$ matrix is diagonal, with
$g_{xx}=g_{yy}=g_{\perp}$ and $g_{zz}=g_{\|}$.
The contribution of the remote band is reverse proportional to the
band distance. Therefore, the closest bands have most significant
effects. However, for $g_{\perp}$, there are no direct corrections
from the closest $L_{3v}$ and $L_{3c}$ bands due to the symmetry.
Instead, the  corrections come from the
indirect ones through the mediation of far-away remote
bands.\cite{baron,roth} With the next closest $L_{2c}$ 
band included,
$g_{\perp}$ reads,\cite{roth}
\begin{eqnarray}
  \nonumber
  g_\perp-g_0&=&\mbox{Re}\frac{4}{m_0i}\sum_{\mu}
  \frac{\langle L_{1c}|p_y|L_{3\mu}\rangle}
  {(E_{L_1c}-E_{L_{3\mu}})(E_{L_{1c}}-E_{L_{2c}})}
  \\
  && \times \langle L_{3{\mu}}|h_x|L_{2c}\rangle\langle
  L_{2c}|p_z|L_{1c}\rangle\ .
  \label{eq4}
\end{eqnarray}
It is expected that this value is small due to the large gap
between $L_{1c}$ and $L_{2c}$ bands.
In order to calculate the contribution of the remote bands 
quantitatively, one needs the
matrix elements of ${\bm h}$ and ${\bm p}$ which are not accessible
theoretically by the $\mathbf{k}\cdot\mathbf{p}$ calculation
alone. In practice, these elements can be obtained by fitting
the experiment data, such as effective masses and band gaps.
In the framework of the $\mathbf{k}\cdot\mathbf{p}$ theory,
the anisotropic effective mass $m^{\ast}$ of $L$-point is\cite{roth}
\begin{equation}
  \frac{m_0}{m^\ast}={\bm 1}+\frac{2}{m_0}\sum_{\mu}\ \!\!^\prime
  \frac{
    \langle L_{1c}| {\bm p} |\mu\rangle
    \langle \mu|{\bm p}|L_{1c}\rangle}{E_{L_{1c}}-E_{\mu}}.
  \label{eq2}
\end{equation}
For the longitudinal mass,
\begin{equation}
  \label{eq:m_l}
  \frac{m_0}{m_l}=
  {m_0\over m_{zz}}=1+\frac{2}{m_0}\frac
  {|\langle L_{1c}|p_z|L_{2c}\rangle|^2}
  {E_{L_{1c}}-E_{L_{2c}}}\ .
\end{equation}
The $L_{3v}$ spin-orbit split off band gap $\delta=2i\langle
L_{3v}|h_z|L_{3v}\rangle$.
By assuming that all the non-vanishing ${\bm p}$ elements in
Eq.~(\ref{eq4}) and (\ref{eq:m_l}) are the same, and all of the
non-vanishing ${\bm h}$ elements are equal to each other,
$g_{\perp}$ can be expressed as
\begin{equation}
  g_\perp-g_0\cong-\delta(m_0/m_l-1)/(E_{L_{1c}}-E_{L_{3v}})\ .
  \label{eq5}
\end{equation}
Using the experiment data of $m_l$ and $\delta$, one
can justify that the correction to $g_{\perp}$ is rather small for
both GaAs and AlAs due to the large
gap between $L_{1c}$ and $L_{3v}$/$L_{2c}$
bands. Take GaAs as an example,
$m_l/m_0=1.9$ while $\delta=0.22$~eV is much smaller than the band
gap (about $3$~eV).\cite{madelung,klimeck} As a result,
$g_{\perp}$ differs from $g_0$ by only 2\%.
The closeness
of $g_{\perp}$ to $g_0$ at $L$-point is not limited to GaAs but rather
universal property of semiconductor with diamond and zinc-blende
structures, in which the symmetry eliminates the direct
correction from the closest bands.\cite{roth}

The parallel component $g_{\|}=g_{zz}$ is very different.
It reads\cite{roth}
\begin{eqnarray}
  \nonumber
g_\|-g_0&=&\mbox{Re}\frac{4}{m_0i}\sum_{\mu\nu}
\frac{\langle L_{1c}|p_x|L_{3\mu}\rangle}
{(E_{L_{1c}}-E_{L_{3\mu}})(E_{L_{1c}}-E_{L_{3\nu}})}
\\ && \times
\langle L_{3\mu}|h_z|L_{3\nu}\rangle
\langle L_{3\nu}|p_y|L_{1c}\rangle\ .
\label{eq7}
\end{eqnarray}
With the
modifications from $L_{3c}$ and $L_{3v}$ bands, it can be
written as\cite{baron}
\begin{equation}
  g_\|-g_0=-\tfrac{2m_0}{\hbar^2}[P^2\tfrac{\delta}{E_g( E_g+\delta)}+
  P^{\prime2}\tfrac{\delta^\prime}{E_g^{\prime}(E_g^{\prime}+\delta^\prime)}].
  \label{eq8}
\end{equation}
Here $E_g=E_{L_{1c}}-E_{L_{3v}}$ and $E_g^\prime=E_{L_{3c}}-E_{L_{1c}}$ are
the band gaps.
$\delta^\prime$ is the spin-orbit splitting of the
$L_{3c}$ conduction band.
$-im_0P(P^{\prime})/\hbar$ is the non-vanishing
inter-band momentum matrix element between $L_{1c}$ and $L_{3v}
(L_{3c})$ bands.
Among these parameters, the band gaps and the
spin-orbit splittings can be measured directly or obtained through
band structure calculations.
In our calculation, we use the experiment data from
Refs.~\onlinecite{madelung} and \onlinecite{klimeck} when
available. For the parameters without  experiment data,
theoretical results from the tight-binding model are
used.\cite{jancu3,fu} Since there is no value of $\delta^\prime$ in the
  literature, we calculate it using the
tight-binding model with the tight-binding
parameters taken from Ref.\  \onlinecite{jancu3}.
All the parameters in Eq.\ (\ref{eq8}) are listed in
Table~\ref{table1}, except $P$ and $P^\prime$. However, there are
neither direct experimental data nor theoretical
results from three band model on $P$ and $P^{\prime}$.
In  $\mathbf{k}\cdot\mathbf{p}$ band structure calculation, both are
determined by fitting other experimentally measurable
parameters. Here we use the transversal effective mass as a
fitting target.
From Eq.~(\ref{eq2}), the transversal effective mass $m_t=m_{xx}$
reads\cite{baron}
\begin{eqnarray}
  \tfrac{m_0}{m_t}-1=\tfrac{m_0}{\hbar^2}
\left[P^2(\tfrac{1}{E_g}+\tfrac{1}{E_g+\delta})
  -P^{\prime2}(\tfrac{1}{E_g^\prime}+\tfrac{1}{E_g^\prime+\delta^\prime})\right].
  \label{eq9}
\end{eqnarray}
Besides this, one more relation, i.e., $P^\prime/P$,
 is required in order to obtain
$P$ and $P^{\prime}$. There are three possible ways
to estimate the value of $\lambda=P^{\prime}/P$.
The first is based on the assumption that
this ratio at the same symmetry point is an universal
constant for different materials.\cite{chadi}
From the experiment data of Ge given in Refs.~\onlinecite{baron} and
\onlinecite{madelung}, i.e.,
$E_{L_{1c}}=0.744$\ eV, $E_{L_{3v}}=-1.53$\ eV,
$E_{L_{3c}}=4.3$\ eV, $\delta=0.23$\ eV, $\delta^\prime=0.27$~eV,
$m_t=0.0791m_0$ and $g_\|=0.82$, we
obtain $\lambda=0.22$ at $L$-point from Eqs.\ (\ref{eq8}) and
(\ref{eq9}).\cite{hermann} With this value, one finds that
$g_{\|}=1.09$ and $1.41$ for GaAs and AlAs respectively. This set of
 parameters are listed in Table~\ref{table1} with rows labeled
``$\alpha$''.
The second way is to assume that
$\lambda$ at different valleys are the same. In this way, we can use
the value of $\lambda$ at $\Gamma$-point where the band structure is
well studied.\cite{winkler}
For GaAs (AlAs), $\lambda=0.456$ ($0.533$), which
give $g_{\|}=0.89$ ($0.93$). These values are
listed in Table~\ref{table1} in the rows labeled ``$\beta$''.
The third way to choose $\lambda$ is to set it to be zero and
thus reduce the three-band model  to two-band one.\cite{baron,roth}
The corresponding results are shown in Table.~\ref{table1} with rows labeled
``$\gamma$''.
From the table, one can see that the remote bands have significant
effect on $g_{\|}$. However, the contribution of the remote bands
strongly depends on how the remote bands are treated.
We comment that all the three estimations of $\lambda$ above
have their reasonings. However, one has no way to judge which one is the
best without experiments.
Fortunately, there is certain important case where the corresponding
$g$-factor does not
depend on the model one uses,
which we address in the following.

From the table, one finds that $g$-factor in $L$-valley is highly
anisotropic. When an
applied magnetic field departs from the
principle axis, the corresponding $g$-factor is a combination of $g_{\perp}$ and
$g_{\|}$. Moreover, different valleys also contribute differently.
The overall $g$-factor should be
averaged over the four $L$-valleys.
However, when the magnetic field is along
some highly symmetric crystal axis, such as $[100]$ direction (i.e., in
Voigt configuration), all the four
$L$-valleys have the same overall $g$-factor. In this case, the
$g$-factor can be expressed as
\begin{equation}
g_{x}=\sqrt
{(g_\perp^2+g_\|^2\mbox{sin}^2\theta+g_\perp^2\mbox{cos}^2\theta)/2},
\label{eq10}
\end{equation}
with $\mbox{cos}^2\theta=1/3$ for $[100]$ direction.  $g_x$ is also
listed in Table.~\ref{table1}.  One finds that the values of $g_x$
obtained from the three different choices of $\lambda$ are very close
to each other.  The difference in $g_x$ given by different models is
less than 5\%, even though the difference in $g_{\|}$ can be as large
as 20-30\%.  The closeness of $g_{x}$ from the three results is not a
coincidence. One can see from Eq.~(\ref{eq10}) that under the Voigt
configuration, the contribution of $g_{\perp}^2$ to $g^2_x$ is more
than twice as that of $g_{\|}^2$. Moreover, the remote bands have
marginal effect on $g_{\perp}$. As a result, $g_x$ is model
insensitive.  This result is of particular interest, since the applied
magnetic field is exactly along the $[100]$ direction in the Voigt
configuration, which is widely used in the Kerr rotation
experiments.\cite{spin} We believe that the obtained $g_x$ is reliable
in studying the spin dynamics in $L$-valley in Voigt configuration.

In summery, we have investigated the $L$-valley Land\'e $g$-factor in bulk
GaAs and AlAs using a three-band
$\mathbf{k}\cdot\mathbf{p}$ model.
The parameters used in the calculation are from experiment data
  if available, or from the tight-binding calculation otherwise.
The $g$-factor in the $L$-valley is highly
anisotropic and can be characterized by two components: the
transversal one $g_{\perp}$, corresponding for the magnetic field
perpendicular to the $L$-axis, and the longitudinal one $g_{\|}$,
for the magnetic field parallel to the $L$-axis. The
transversal component is close to the free electron $g$-factor due to
symmetry. Whereas the longitudinal component is shown to be strongly affected by
the remote bands.
The contribution of the remote bands depends on how these
bands are treated, and the results are quite different.
It is hard to judge which one is more reasonable without the
justification of the experiments. However, when the magnetic field is
in Voigt configuration (i.e.,
along $[100]$ axis) which is
widely used in  experiments investigating spin dynamics,
different methods used in this investigation give almost
identical  Land\'e factors: $g_x\simeq 1.77$ for GaAs and
$1.85$ for AlAs.

This work was supported by the Natural
Science Foundation of China under Grant Nos.\ 10574120 and 10725417, the
National Basic Research Program of China under Grant
No.\ 2006CB922005 and the Knowledge Innovation Project of Chinese Academy
of Sciences.

\end{document}